\documentclass{JHEP}
\def\beqra{\begin{eqnarray}}
\def\eeqra{\end{eqnarray}}
\def\beq{\begin{equation}}
\def\eeq{\end{equation}}
\def\ds{\displaystyle}

    \def\L{\Lambda}

    \def\T{\Theta}

\def\half{\mbox{\small$\frac{1}{2}$}}

\def\lta{\mathrel{\vcenter{\hbox{$<$}\nointerlineskip\hbox{$\sim$}}}}

\newcommand{\Tr}{\mathop{\mathrm{Tr}}}

\input  epsf.sty
\title{Wilsonian Renormalization Group\\
and the Non-Commutative IR/UV Connection}
\author{Luca Griguolo \\Dipartimento di  Fisica, Universit\`a  di Parma, 

INFN-Gruppo Collegato di Parma

Parco Area delle Scienza 7/A, 43100 Parma, Italy\\ 
E-mail: \email{griguolo@fis.unipr.it}}

\author{ Massimo Pietroni

 \\ INFN - Sezione di Padova,

 Via F. Marzolo 8,

I-35131   Padova,  Italy \\ E-mail: \email{pietroni@pd.infn.it}}

 \date{data}

  \maketitle

\abstract{We study the IR/UV connection of the four-dimensional 
non-commutative $\phi^4$ 
theory by using the Wilsonian Renormalization Group equation. 
Extending the usual 
formulation to the non-commutative case we are able to prove  UV 
renormalizability to all orders in perturbation theory. The full RG equations 
are finite in the IR, but  perturbative approximations of them are plagued 
by IR divergences. The latter can be systematically resummed in a way 
analogous to what is done in finite temperature field theory. 
As an application, next-to-leading order corrections to the two-point 
function are 
explicitly computed. The usual Wilsonian picture, {\it i.e.} 
the insensitivity of 
the IR regime to the UV, does not hold in the non-commutative case. 
Nevertheless
it can be 
partially recovered by a matching procedure, in which a high-energy theory, 
defined in the deep non-commutative regime, is connected at some intermediate 
scale to a commutative low-energy theory. The latter knows about 
non-commutativity only through the boundary conditions for two would-be 
irrelevant couplings.} 

\preprint{DFPD 01/TH/15\\
UPRF-2001-09}

\begin{document} 

\section{Introduction} 
\label{INTRO}
Recently a lot of interest has been devoted to the study of quantum field 
theories on non-commutative spaces. The main  motivation arises 
directly from their tight relation with string theories: low energy 
excitations of a $D$-brane in a magnetic $B_{\mu\nu}$ background are indeed 
described by field theories with space non-commutativity \cite{seiwit}. In 
this limit the relevant description of dynamics is in terms of massless 
open string states, while massive open string states and closed strings 
decouple: the full consistent string theory seems therefore truncated to the 
usual field theoretical degrees of freedom. Explicit computations have been 
performed in \cite{russo}, showing that the robustness of the above 
picture holds even after string loop effects are included: 
this strongly suggests the 
possibility that the related quantum field theories are well defined too. 
On the other hand the consistency of the latter is far from being obvious 
when examined 
from a purely field theoretical point of view: they are non-local 
(involving an arbitrarily high number of derivatives in the couplings) and 
there is a new dimensional parameter, other than the masses, taking into 
account the scale at which non-commutativity becomes relevant. 
In particular, unitarity and renormalizability may be in jeopardy: 
it was shown in fact in \cite{gomi,alva} that when the non-commutativity 
involves space {\em and} time the perturbative unitarity 
is in trouble, while in the pure spatial case 
 consistency with the 
Cutkoski's rules and positivity properties has been checked 
\cite{gomi,alva}. 
While these results could be expected from string theory -- as massive 
string states do not decouple in the 
 case of space-time non-commutativity \cite{om} --  the issue of 
renormalizability is more subtle. Contrary to 
early suggestions it was in fact found \cite{filk} that infinities appear 
when perturbative computations are performed in non-commutative 
scalar theories, the extension of this result to fermionic, gauge and 
supersymmetric theories being straightforward. The non-local character 
of the theory could therefore invalidate the usual proofs of 
renormalizability, which are based on the polynomial nature of the 
divergent terms in perturbation theory. Moreover, an highly non-trivial
 mixture 
between ultraviolet (UV) and infrared (IR) phenomena was observed
first in scalar 
theories \cite{minwa} and then in gauge theories \cite{suski}: 
this property, being probably the most surprising feature of non-commutative 
quantum field theories, is known as the IR/UV mixing. It happens that some UV 
divergences are  regulated by an effective cut-off $O(M_{NC}^2/p)$, where 
$M_{NC}$ is the typical scale induced by non-commutativity and $p$ a typical 
external momentum. Then,  as $p \to 0$, a singular IR behavior      
appears in the perturbative results, even for massive theories, which is 
actually the remnant of UV divergences of the corresponding 
commutative theory. 
If $\Lambda_0$ is the overall UV cut-off, it is easy to show that 
the limit $p \to 0$ does not commute with $\Lambda_0\to\infty$, suggesting a 
dangerous dependence of infrared physics from very high-energy modes. The 
basis of the program of renormalization seems therefore in trouble and 
doubts have been casted on the predictivity of the theory 
itself. Moreover, 
the IR behavior becomes extremely problematic when 
the  IR/UV terms appear as subdiagrams at higher orders. For instance, 
in the massless theory
the would-be one-loop UV divergence, regulated by the non-commutativity, 
finds a new avatar as a 
two-loop IR divergence, independently of the scale of the external momenta. 

Concerning the UV behavior, the only four-dimensional theory that has been 
proved to be renormalizable at any loop order 
is the Wess-Zumino model \cite{giro}. 
Actually, supersymmetric theories are easier to investigate because the 
IR/UV mixing generates only logarithmic divergences in the external momenta 
\cite{suski} (this is related to the absence of quadratic 
divergences in their commutative counterpart). Investigations of 
non-commutative supersymmetric Yang-Mills were presented in \cite{vale}.
No general result in 
this sense has been available up to now for scalar or gauge theories beyond 
two-loops, mostly due to the extreme complexity of the non-commutative 
diagrammatics when higher orders are involved. Even in the simple case of
the (non-commutative) $\phi^4$ theory
convergence 
theorems and recursive subtractions fail to give a definitive 
answer to the question \cite{roiban}, 
although renormalizability has been argued since 
the original quantum investigation \cite{minwa}. Explicit two-loop 
computations have been performed in the massive case, 
showing that the theory  can be effectively renormalized at this order: 
at the same time  it was noticed  that at higher loops problems arise, 
due to divergences induced by the presence of the above mentioned IR/UV terms
\cite{belov}. Non supersymmetric gauge theories have been discussed so far 
at one-loop \cite{gauge}. Actually, in ref.~\cite{pacchi} 
it has been claimed that 
non-commutative QED is renormalizable at all orders. 
However, that paper is based on  mapping  the non-commutative
gauge theory on a commutative one, where the IR/UV problem is absent. The 
connection between this approach and the usual diagrammatic one is not clear.

The purpose of this paper is to study the IR/UV connection of the 
non-commutative $\phi^4$ theory using the Wilsonian Renormalization Group 
equation (RG) \cite{willi}, as formulated by Polchinski \cite{polchinski} in
the case of quantum field theories. The RG turns out to be a very powerful tool
in order to disentangle the IR side of the problem from the UV one.
The main
feature that we will use is the introduction of an explicit momentum cut-off,
$\L$, which can take any value between the UV cut-off,  $\L_0$,
 and zero. The RG 
equations describe the evolution of the couplings of the theory as loops
with momenta between $\L_0$ and $\L$ are included, with $\L$ eventually
going to the physical value $\L\to 0$. 
A two-steps strategy is then possible. First, take $\L$ much larger
than any physical mass scale (but $\L\ll\L_0$) and sample the UV sector of the
theory,
discussing under which conditions the $\L_0\to \infty$ limit can be taken
(UV renormalizability). Second, study the IR limit by sending $\L\to 0$ (IR
finiteness).

Besides the neat separation between UV and IR, there is another feature
of the RG which will be of great use in the following. The RG equations are 
formally one-loop equations in which  the tree-level vertices and propagators
are replaced by full-- $\L$-dependent-- ones. Thus, perturbation theory is 
reproduced by solving the full equations iteratively, putting the $n$-th 
order propagators and vertices in and getting those at $(n+1)$-th order
out  upon integration.

The two features above were used in \cite{polchinski} and in \cite{bdm} to 
prove perturbative UV renormalizability of the commutative $\phi^4$ theory
to all orders and, in the massless case, the IR finiteness of the theory
\cite{bdm}. 
These proofs are  extremely simple. They are
essentially based on power counting and do not require any analysis of 
overlapping divergences of Feynman diagrams. As we will see, this
holds for the non-commutative case as well. 

Our results can be summarized as follows:
\begin{itemize}
\item[{i)}] UV renormalization is proved to all orders in perturbation theory;
\item[{ii)}] the full RG equations are finite in the IR, but perturbative
approximations of them are plagued by IR divergences. The latter can be
systematically resummed in a way analogous to  what is done at finite 
temperature (Hard Thermal Loop (HTL) resummation \cite{bp,parwani}). 
Next-to-leading order corrections to the two-point function 
are explicitly calculated (see also \cite{gp1});
\item[{iii)}] the  usual Wilsonian picture, {\em i.e.} the insensitivity of 
the IR regime to the UV, is lost in the non-commutative case. Nevertheless,
it can be partially recovered by a matching procedure, in which a high-energy 
theory, defined in the deep non-commutative regime, is connected at some 
intermediate scale to a commutative low-energy theory. The latter knows
about high-energy non-commutativity only via the boundary conditions for
two would-be irrelevant couplings.
\end{itemize}

The plan of the paper is the following. In sect.~\ref{NCQFT} we introduce the 
non-commutative version of the four-dimensional real $\phi^4$ theory, 
we display the Feynman rules and we discuss the 
difference between planar and non-planar diagrams in a specific example, 
the two-point function. The IR/UV mixing is presented by a one-loop 
computation and the problems arising when higher orders are 
considered are exemplified. In sect.~\ref{BDM} we review the Wilsonian RG 
approach to UV renormalization and IR finiteness in the commutative case, 
illustrating the 
procedure developed in \cite{bdm}. In sect.~\ref{NCWRG} we extend the 
formulation of the RG to the non-commutative case and
we prove the UV renormalizability of the theory to all perturbative orders.
Then, we discuss the IR regime. We show the necessity of a resummation
and perform it by modifying the tree-level propagator and adding a 
proper counterterm in order to avoid overcounting. As an application, 
we review our recent 
calculation of next-to-leading corrections to the two-point function 
\cite{gp1}. 
In sect.~\ref{PRED} we discuss the sensitivity of the theory in the 
IR regime to the high-energy theory. We point out that UV/IR mixing destroys 
the usual Wilsonian picture and discuss how it can be partially recovered
by a suitable matching procedure.
Finally, in sect.~\ref{CONC} we summarize our results and discuss some possible
implications of them. 
\section{Non-commutative field theory}
\label{NCQFT}
Non-commutative spaces are defined by the following generalization of the
usual quantum commutation relations 
\beq
\label{COMM}
[\hat{x}^\mu,\hat{x}^\nu]=i \T^{\mu \nu}\,,\,\,\,\,\,\,[\hat{x}_i,\hat{p}_j]=
i \delta_{i j}\,,
\eeq
where $\T^{\mu \nu}$ is a constant anti-symmetric matrix of dimension 
$[M]^{-2}$, greek indices run from zero to three and latin ones from one 
to three. It was shown in ref.~\cite{gomi,alva} that when any $\T^{0 i}$ is 
different from zero perturbative unitarity is in trouble, so that in order
to get a consistent field theory one should consider only spatial indices 
also in the first of eqs.~(\ref{COMM}). 
Since we will work in Euclidean space, we will ignore the problem and consider 
a generic $\T$-matrix. 
 
A non-zero $\T^{\mu \nu}$ breaks Lorentz invariance, therefore 
explicit dependence on \mbox{$\tilde{p}^\mu \equiv \T^{\mu\nu}p^\nu$}, 
with $p^\nu$ a generic
external momentum, is to be expected on general grounds.

In order to construct a field theory, it is convenient to use 
the so-called Weyl-Moyal correspondence 
%cite{gaume}, 
which amounts to work 
on
the usual commutative space while redefining the multiplication between
 functions
according to
\beq
\label{STAR}
f_1(x) f_2(x) \rightarrow 
(f_1 \star f_2)(x)\equiv \left. e^{\frac{i}{2}\T^{\mu\nu}
\partial^y_\mu \partial^z_\nu}f_1(y) f_2(z)\right|_{y=z=x}\,.
\eeq
It is easy to check that the Moyal bracket between the commuting 
$x_\mu$ and $x_\nu$ consistently replaces the relations in (\ref{COMM}),
that is 
\[
[x^\mu,x^\nu]_{MB} = x^\mu \star x^\nu - x^\nu \star x^\mu = i \T^{\mu \nu}\,.
\]
The action for the Euclidean real scalar theory becomes
\label{ACTION}
\beqra
&& S[\phi]= \frac{1}{2}\ds\int\frac{d^4 p}{(2 \pi)^4} \left[ \phi(p) (p^2+m^2) 
\phi(-p)\right] \nonumber \\
&& \ds +
\frac{g^2}{4!} \int \left(\prod_{a=1}^4 \frac{d^4 p_a}{(2 \pi)^4}
  \phi(p_a)\right) (2\pi)^4\delta^{(4)}\left(\sum_{a=1}^4 p_a \right)
e^{i p_{12}} e^{i p_{34}}\,,
\eeqra
where
\[p_{ab}\equiv \frac{p_a \cdot \tilde{p}_b}{2}\;.\]
As we see, the quadratic part is the same as in the commutative case, due
to the antisymmetry of $\T^{\mu\nu}$, so the propagator is still 
 $ D(p)=(p^2+m^2)^{-1}$, while the Feynman rule for the interaction vertex gets
modified according to
\beqra
\label{FRULES}
\ds 
&&\Gamma^{(4)}(p_1,p_2,p_3,p_4)= g^2 h(p_1,p_2,p_3,p_4) \nonumber\\
&&\equiv 
\frac{g^2}{3}\left[\cos(p_{12})\cos(p_{34})
+ \cos(p_{13})\cos(p_{24})
+\cos(p_{14})\cos(p_{23})
\right]\,.
\eeqra
The Moyal phase factor
modifies the behavior of loop integrals with respect to the commutative
case. Consider the one-loop correction to the self-energy
\beqra
\label{SE1LOOP}
\Sigma_{\L_0}(p)&&=
-\ds \frac{g^2}{6}\int\frac{d^4q}{(2\pi)^4} \Theta(\L_0^2-q^2) 
\frac{1}{q^2+m^2}\left[2+\cos( q\cdot\tilde{p})\right] \nonumber\\
&&=-\frac{g^2}{32 \pi^2}\left[\frac{2}{3}\L_0^2 + \frac{4}{3~ \tilde{p}^2 }
\left(1-J_0(\L_0~\tilde{p})\right)\right]\;,\;\;\;(m^2 \ll 
\mathrm{min}\{\L_0^2,1/\tilde{p}^2\})
\eeqra
where we have regulated
the UV quadratic divergences by means of a momentum cut-off. The first
term in parenthesis exhibits the usual $\L_0^2$ behavior of the commutative case 
with a coefficient $2/3$ instead of $1$. This is the contribution of the 
so-called `planar' graphs, in which two nearby scalar legs are contracted in
order to get the tadpole loop. The second term comes from the 
non-planar graphs and contains a non-vanishing phase factor. 
It exhibits the two limiting behaviors 
\beqra
\label{LIMITS}
\ds \frac{4}{3~ \tilde{p}^2 }
\left(1-J_0(\L_0~\tilde{p})\right) &&  \to \frac{1}{3} \L_0^2
\;\;\;\;\;\;\;\;\;({\mathrm as}~\L_0~\tilde{p}\to 0)\nonumber\\
&&
\to \frac{4}{3}\frac{1}{\tilde{p}^2} 
\;\;\;\;\;\;\;\;\;({\mathrm as}~\L_0~\tilde{p}\to \infty)\;,
\eeqra
from which we see that if we take the commutative limit $\T^{\mu\nu} \to 0$ 
before removing the cut-off, the `missing' UV divergence $ \L_0^2/3$ is 
recovered. On the other hand, if we let $\L_0\to \infty$ first, the 
oscillating term provides an effective UV cut-off $2/|\tilde{p}|$, which
regulates the integral leaving a $\L_0$-independent term.
Thus, in the non-commutative theory we generally have:
\begin{itemize}
\item[i)] different coefficients for the UV divergent terms;
\item[ii)] new {\em IR-dangerous terms}, induced by the 
{\em effective UV cut-off}, 
$2/|\tilde{p}|$.
\end{itemize}
It is rather clear that when we insert the tadpole (\ref{SE1LOOP}) in 
higher order contributions to the self-energy, we get more and more IR 
divergent integrals (we present only the leading contribution to the 
singularity at a given order)
\beq
\label{IRPROB}
g^2 \int\frac{d^4q}{(2\pi)^4}\frac{1}{(q^2+m^2)^{n+1}} \Sigma(q)^n \to 
g^{2n+2} \int\frac{d^4q}{(2\pi)^4}\frac{1}{(q^2+m^2)^{n+1}} 
\frac{1}{\tilde{q}^{2n}} \;\;\;\;
(\mathrm{IR})\;,
\eeq
which make the self-energy IR divergent at $O(g^4)$ in the massless ($m=0$) 
case and at $O(g^6)$ in the massive one.

The connection between the UV and IR divergences that we have just outlined 
makes a proof of perturbative renormalizability along the usual lines
quite cumbersome. The main difficulty lies manifestly in the difficulty in
disentangling the UV from the IR sectors of the loop integrals.
Indeed, the possibility of absorbing UV divergences by means of
local counterterms has been discussed at two-loop in 
refs.~\cite{roiban,belov,micu}, but no finite result could be obtained at 
that order, in the massless theory, due to the pathological behavior of the 
integrals in the IR. Although this difficulty was recognized since the 
original work on the subject \cite{minwa}, no proposal has been done  
up to now to systematically handle these divergences and 
no results have been presented by taking consistently into account 
higher orders.

In the following we will show that the use of Wilsonian methods, 
where an explicit
momentum cut-off separates the IR from the UV, is  of great help in 
organizing a perturbative proof of renormalizability. At the same time, 
the structure of the exact RG equations suggests an 
appropriate procedure to tame IR singularities.

%FOOTNOTE\footnote{
%In ref.  $\L$ was kept finite in eq.~(\ref{LIMITS})
%so that the 
%two-loop correction was claimed to have a safe IR behavior.
% Indeed in 
%this case the integral in eq.~(\ref{IRPROB}) receives a contribution
%$O(\L^2 \log(m^2))$ from the IR region, which is manifestly divergent when
%the $\L\rightarrow \infty$ limit is eventually taken.}

\section{Wilsonian RG in the commutative case}
\label{BDM}
In this section we review the formulation of the RG \`a la 
Polchinski \cite{polchinski} in the commutative theory. Following 
ref.~\cite{bdm}, we will review the use of the RG to demonstrate
perturbative renormalizability of the real scalar theory 
and IR finiteness in the massless case.
\subsection{The Wilsonian flow}
Our starting point is the path integral
\beqra
\label{Z}
\ds Z[J]=e^{W[J]}=\int {\cal D} \phi \:\:&& \exp\left\{-\frac{1}{2} 
\int\frac{d^4p}{(2\pi)^4} \phi(p) D^{-1} \phi(-p) - S_{\mathrm int}[\phi] 
\right.
\nonumber\\
&& \:\:\:\:\:\ds +\left.
\int\frac{d^4p}{(2\pi)^4} \phi(p)J(-p)\right\}\;,
\eeqra
where the interaction action $ S_{\mathrm int}[\phi]$
contains the 
bare couplings and has a $Z_2$ $\phi\to -\phi$ symmetry.
%\beq
%\label{SINT}
% S_{\mathrm int}[\phi]=\half \int\frac{d^4p}{(2\pi)^4} \phi(p)
%\left[\gamma_2^{(B)}
%(p^2-\mu^2)+ \gamma_3^{(B)}\right]\phi(-p) + 
%\frac{\gamma_4^{(B)}}{4 !} \int d^4 x \:\Phi(x)^4\;,
%\eeq
%contains the bare `relevant' couplings $\gamma^{(B)}_{2,3,4}$ which are fixed
%by imposing the renormalization conditions on the full two-- and four--point
%functions,
%\beq
%\label{RCOND}
%\Gamma_2(p^2=\mu^2)=\mu^2,\;\;\;\;\;\left.
%\frac{d \Gamma_2(p^2)}{d p^2}\right|_{p^2=\mu^2}= 1,\;\;\;\;\;
%\Gamma_4(\{\bar{p}_i\})= g^2\;,
%\eeq
%with $\bar{p}_i\cdot \bar{p}_j=\mu^2 (\delta_{ij}-\frac{1}{4})$.
We now introduce an UV cut-off, $\L_0$, and a IR one, $\L$, by making 
the substitution
\[
D(p)\rightarrow D_{\L, \L_0}(p) \equiv D(p) K_{\L, \L_0}(p)
\] in eq.~(\ref{Z}), where $K_{\L, \L_0}(p)$ is equal to one for 
$\L^2<p^2<\L_0^2$ and vanishes rapidly outside.
The substitution above defines $Z_{\L, \L_0}$ and $W_{\L, \L_0}$, the generating
functionals of Green functions in which only momenta between $\L$ and $\L_0$ 
have been integrated out. By differentiating w.r.t $\L$ we get the RG 
equation for $Z_{\L, \L_0}$
\beq
\label{RGMUM}
\L\frac{\partial\;\;}{\partial \L} Z_{\L, \L_0}[J] =-\half 
\int\frac{d^4 q}{(2\pi)^4} 
\L\frac{\partial\;\;}{\partial \L} D_{\L, \L_0}^{-1}(q) (2 \pi)^8
\frac{\delta^2 Z_{\L, \L_0}[J] }{\delta J(q)\delta J(-q)}\;.
\eeq
In order to discuss the issue of renormalizability, it is more convenient to 
consider
the 1PI generating functional, defined as usual as
\beq
\label{1PI} \Gamma_{\L, \L_0}[\phi]= -W_{\L, \L_0}[J] + W_{\L, \L_0}[0] + 
\int\frac{d^4 p}{(2\pi)^4}\phi(p)J(-p)\;,
\eeq
where $\ds \phi(p)=\frac{1}{(2 \pi)^4}
\frac{\delta W_{\L, \L_0}[J]}{\delta J(-p)}$.

Cut-off, 1PI,  2n-point functions are then given by 
\beq
\label{NPOINT}
(2 \pi)^4 \delta^{(4)}(\sum_i p_i) 
\Gamma^{2 n}_{\L, \L_0}(p_1,\ldots,p_{2 n}) = 
(2 \pi)^8 \frac{\delta^{2 n} \Gamma_{\L, \L_0}[J]}{\delta \phi(p_1)\cdots
\delta \phi(p_{2 n})}\;.
\eeq
From the definitions above and from the RG equation (\ref{RGMUM}) we get
the evolution equations for the 1PI Green functions \cite{bdm}. Isolating
the interacting part of the two-point
function,
\beq
\label{2PF}
\Gamma^2_{\L, \L_0}(p)= D^{-1}_{\L, \L_0}(p)+ \Sigma_{\L, \L_0}(p)\;,
\eeq
we have
\beq
\label{RGE2p}
\ds \L\frac{\partial\;\;}{\partial \L}\Sigma_{\L, \L_0}(p) = \half
\int\frac{d^4 q}{(2\pi)^4}\frac{S_{\L, \L_0}(q)}{q^2+m^2} 
\Gamma^4_{\L, \L_0}(q,p,-p,-q)\;,
\eeq
while for  $n \ge 2$ we have 
\beq
\label{RGEnp}
 \L\frac{\partial\;\;}{\partial \L} \Gamma^{2n}_{\L, \L_0}(p_1,\ldots,
p_{2n}) = 
\half
\int\frac{d^4 q}{(2\pi)^4}\frac{S_{\L, \L_0}(q)}{q^2+m^2} 
\overline{\Gamma}^{2n+2}_{\L, \L_0}(q,p_1,\ldots,
p_{2n},-q)\;,
\eeq 
with
\beqra
\label{GBAR}
\ds 
&& \overline{\Gamma}^{2n+2}_{\L, \L_0}(q,p_1,\ldots,
p_{2n},q^\prime)=\Gamma^{2n+2}_{\L, \L_0}(q,p_1,\ldots,
p_{2n},q^\prime)- \nonumber \\
\ds && \sum_{k=1}^{n-1} 
\Gamma^{2k+2}_{\L, \L_0}(q,p_{i_1},\ldots,
p_{i_{2k}},-Q) \frac{1}{\Gamma^{2}_{\L, \L_0}(Q)}
\overline{\Gamma}^{2n-2k+2}_{\L, \L_0}(Q,p_{i_{2k+1}},\ldots,
p_{i_{2n}},q^\prime)\;.
\eeqra
The kernel in eqs.~(\ref{RGE2p},\ref{RGEnp}) is given by
\beqra
\label{KERNEL}
\ds \frac{S_{\L, \L_0}(q)}{q^2+m^2} 
&& \ds \equiv
\L\frac{\partial\;\;}
{\partial \L} 
\left.\frac{1}{(q^2+m^2) K_{\L,\L_0}(q)^{-1}+\Sigma_{\L^\prime,\L_0}(q)}
\right|_{\L^\prime=\L}\,,\nonumber\\
&&\ds 
=\frac{1}{q^2+m^2} 
\frac{1}{\left[ 1+\frac{\Sigma_{\L, \L_0}(q)}{q^2+m^2}
K_{\L,\L_0}(q)  \right]^2} \L\frac{\partial\;\;}
{\partial \L}K_{\L,\L_0}(q)\;.
\eeqra
There is a simple recipe for deriving the RG equation for a given
Green function:
\begin{itemize}
\item[{ i)}] write the one-loop expression for $\Gamma^{(2n)}$ 
obtained by using
all the vertices up to $\Gamma^{(2n+2)}$, as if they were formally tree-level;
\item[{ ii)}] promote  the tree-level 
vertices above to full, running, vertices,
$ \Gamma^{(2n)}\rightarrow \Gamma_{\L,\L_0}^{(2n)}$, and the tree-level 
propagator to the full, cut-off, propagator;
\item[{ iii)}] take the derivative with respect to $\L$ everywhere in the 
$K$'s but not in the $\Sigma$'s or $\Gamma$'s. 
\end{itemize}
The fact that the RG equations are formally one-loop is crucial  in 
allowing a iterative proof of perturbative renormalizability, as we review in 
the following subsection.

\subsection{RG flows for relevant and irrelevant couplings}
\label{UVREN}
We now impose the renormalization conditions
\beq
\label{RCOND}
\Gamma_2(p^2=p_0^2)=m^2+p_0^2,\;\;\;\;\;\left.
\frac{d \Gamma_2(p^2)}{d p^2}\right|_{p^2=p_0^2}= 1,\;\;\;\;\;
\Gamma_4(\bar{p}_1,\ldots,\bar{p}_4)= g^2\;,
\eeq
with the symmetric renormalization point  defined 
as $\bar{p}_i\cdot \bar{p}_j=p_0^2 (\delta_{ij}-\frac{1}{4})$.
In order to study UV renormalization, the `relevant' operators, {\em i.e.} 
those with non-negative mass dimension, have to be isolated. They are
\beq
\label{RELOPS}
\gamma_2(\L)\equiv 
\left.\frac{d \Sigma_{\L, \L_0}(p)}{d p^2}\right|_{p^2=p_0^2},\;\;\;\;
\gamma_3(\L)\equiv \left.\Sigma_{\L, \L_0}(p)\right|_{p^2=p_0^2},\;\;\;\;
\gamma_4(\L)\equiv 
\Gamma^{4}_{\L, \L_0}(\bar{p}_1,\ldots,
\bar{p}_{4})\,.
\eeq
The two-- and four--point functions can be rewritten as
\beqra
\label{DELTA24}
 &&\ds \Sigma_{\L, \L_0}(p) = \gamma_3(\L) + (p^2-p_0^2) \gamma_2(\L)+
\Delta^2_{\L, \L_0}(p) \nonumber \\
&& \ds \Gamma^{4}_{\L, \L_0}(p_1,\ldots,
p_{4})=\gamma_4(\L) +\Delta^4_{\L, \L_0}(p_1,\ldots,
p_{4})
\eeqra
where $\Delta^2$ and $\Delta^4$ satisfy $\Delta^2_{\L, \L_0}(p^2=p_0^2)=0$,
$\Delta^4_{\L, \L_0}(\bar{p}_1,\ldots,\bar{p}_{4})=0$. They
 are `irrelevant' operators together with all $\Gamma^{2n}_{\L, \L_0}$'s
 with $n>2$. 
The RG equations for relevant and 
irrelevant operators can be read from eqs.~(\ref{RGE2p}),~(\ref{RGEnp}), and
form a system of coupled differential equations giving the evolution of the 
couplings as the IR cut-off $\L$ is lowered from $\L=\L_0$ to $\L=0$. 
The renormalization conditions corresponding to eq.~(\ref{RCOND}) 
are imposed by fixing the boundary conditions for the relevant couplings 
at the physical point $\L=0$:
\beq
\label{BCREL}
\gamma_2(0)=0,\,\,\,\gamma_3(0)=0,\,\,\,\gamma_4(0)=g^2\;.
\eeq
The irrelevant couplings are fixed at the UV point $\L_0$, where they can be
taken to be vanishing
\beq
\label{BCIRREL}
\Delta^2_{\L_0, \L_0}(p)=\Delta^4_{\L_0, \L_0}(p_1,\ldots,
p_{4})=\Gamma^{2n}_{\L_0, \L_0}(p_1,\ldots,p_{2n})=0\;\;\;\;\;\;\;\;\;\;\;
(n > 2)\;.
\eeq
Integrating the RG equations with the above boundary conditions leads to
a set of coupled integral equations 
\beqra
\label{INTEQREL}
\gamma_2(\L) &=& \ds\half \int\frac{d^4 q}{(2\pi)^4}\int_0^\L
\frac{d\lambda}{\lambda} \frac{S_{\lambda, \L_0}(q)}{q^2+m^2} 
\left.\frac{\partial\;\;}{\partial p^2}
\Gamma^4_{\lambda, \L_0}(q,p,-p,-q)\right|_{p^2=p_0^2}\;,\nonumber\\
&&\nonumber\\
\gamma_3(\L) &=& \ds \half \int\frac{d^4 q}{(2\pi)^4}\int_0^\L
\frac{d\lambda}{\lambda} \frac{S_{\lambda, \L_0}(q)}{q^2+m^2} 
\left.
\Gamma^4_{\lambda, \L_0}(q,p,-p,-q)\right|_{p^2=p_0^2}\;,\nonumber\\
&&\nonumber\\
\gamma_4(\L) &=& \ds g^2 + \half \int\frac{d^4 q}{(2\pi)^4}\int_0^\L
\frac{d\lambda}{\lambda} \frac{S_{\lambda, \L_0}(q)}{q^2+m^2} 
\;\overline{\Gamma}^6_{\lambda, \L_0}(q,\bar{p}_1,\ldots,\bar{p}_4,-q)\;,
\eeqra
for the relevant couplings, and
\beqra
\label{INTEQIRREL}
&&\Delta^2_{\L,\L_0}(p) = \ds -\half \int\frac{d^4 q}{(2\pi)^4}\int_\L^{\L_0}
\frac{d\lambda}{\lambda} \frac{S_{\lambda, \L_0}(q)}{q^2+m^2} 
\Delta\Gamma^4_{\lambda, \L_0}(q,p,-p,-q)\;,\nonumber\\
&&\nonumber\\
&&\Delta^4_{\L,\L_0}(p_1,\ldots,p_4) = \ds -\half \int\frac{d^4 q}{(2\pi)^4}
\int_\L^{\L_0}
\frac{d\lambda}{\lambda} \frac{S_{\lambda, \L_0}(q)}{q^2+m^2} 
\Delta\overline{\Gamma}^6_{\lambda, \L_0}(q,p_1,\ldots,p_4,-q)\;,\nonumber\\
&&\nonumber\\
&&\ds \Gamma^{2n}_{\L,\L_0}(p_1,\ldots,p_{2n}) 
= - \half \int\ds\frac{d^4 q}{(2\pi)^4}\int_\L^{\L_0}
\frac{d\lambda}{\lambda} \frac{S_{\lambda, \L_0}(q)}{q^2+m^2} 
\;\overline{\Gamma}^{2n+2}_{\lambda, \L_0}
(q,\bar{p}_1,\ldots,\bar{p}_{2n},-q)\;,
\;\;\;\;(n>2)\nonumber\\
&&
\eeqra
\newpage
where
\beqra
\label{SUBTR}
&& \Delta\Gamma^4_{\lambda, \L_0}(q,p,-p,-q) \equiv 
\Gamma^4_{\lambda, \L_0}(q,p,-p,-q)-
\left.\Gamma^4_{\lambda, \L_0}(q,p,-p,-q)\right|_{p^2=p_0^2}\nonumber\\
&& \;\;\;\;\;\;\;\;\;\;\;\;\;\;\;\;\;\;\;\;\;\;\;\;\;\;\;\;\;\;\;\;\;\;\;\;\ds 
- (p^2-p_0^2)
\left.\frac{\partial\;\;}{\partial p^2}\Gamma^4_{\lambda, \L_0}(q,p,-p,-q)
\right|_{p^2=p_0^2}\;,
\nonumber\\
&&\nonumber\\
&&\Delta\overline{\Gamma}^6_{\lambda, \L_0}(q,p_1,\ldots,p_4,-q) \equiv 
\overline{\Gamma}^6_{\lambda, \L_0}(q,p_1,\ldots,p_4,-q)-
\overline{\Gamma}^6_{\lambda, \L_0}(q,\bar{p}_1,\ldots,\bar{p}_4,-q)\;.
\nonumber\\
&&
\eeqra
The first aspect to notice about the above equations is the different role
 played 
by the cut-off $\L$ for the relevant and the irrelevant operators. In the 
former case, it 
acts as a UV cut-off (eq.~(\ref{INTEQREL})), thus 
ensuring-- by construction-- the finiteness
of the integrals when the UV cut-off $\L_0$ is removed. 
For the irrelevant couplings $\L$ acts instead as a 
IR cut-off. The proof of UV renormalizability then amounts to show that in the 
$\L_0\to\infty$ limit the integrals in eq.~(\ref{INTEQIRREL}) are finite, which is
ensured by dimensional reasons and by the subtractions in eq.~(\ref{SUBTR}).
Moreover, in this framework, the discussion of the behavior in the IR is disentangled by that
in the UV, as we can take the two limits $\L_0 \to \infty$ and $\L\to 0$ at different stages.
This will turn out to provide useful insights to the IR/UV connection in the 
non-commutative theory.

The second relevant aspect of eqs.~(\ref{INTEQREL},\ref{INTEQIRREL}) is that they are 
formally one-loop integrals with loop momentum $q^2\sim \lambda^2$. 
Thus, perturbation theory can be 
reconstructed from the RG equations by putting the $l$-loop result 
on the RHS of eqs.~(\ref{INTEQREL},\ref{INTEQIRREL}) and getting the 
$l+1$-loop result upon 
integration on $\lambda$.

From now on we will use a sharp momentum IR cut-off, {\em i.e.} the kernel
$K_{\L,\L_0}(q)$ of eq.~(\ref{KERNEL}) will contain a 
Heaviside function $\Theta(\sqrt{q^2}-\L)$. Our conclusions on
 renormalizability and
IR finiteness do not rely on the 
type of  cut-off function, but we must specify it in order to get explicit 
results for $\L\neq 0$ and at any finite order in the approximations.
Moreover, the explicit form of the UV cut-off needs not to be specified.
The kernel (\ref{KERNEL}) then reads
\beq
\label{KERNELSHARP}
\frac{S_{\L,\L_0(q)}}{q^2+m^2}=-\frac{\L}{\L^2+m^2}
\delta(\L-\sqrt{q^2}) s(\L)\,,
\;\;\;\;\;\;s(\L)=\left(1+
\frac{\Sigma_{\L,\L_0}(\L)}{\L^2+m^2}\right)^{-2}\,,
\eeq
where the two-point function $\Sigma_{\L,\L_0}(q)$ is evaluated at $q=\L$,
a fact which will turn out to be crucial in the following.

\subsection{Proof of UV renormalizability}
\label{RENCOMM}
In order to simplify the power-counting, we take $\L$ much larger than any
physical scale in the theory, {\em i.e.} $p_0,m\ll \L \ll \L_0$. In this 
limit  the one-loop contributions are given by,
\beqra
\label{SCAL1LOOP}
&&\gamma_2(\L)=0 ,\;\;\;\;\;\;\gamma_3(\L) =-\frac{g^2}{32 \pi^2}\L^2,
\;\;\;\;\;
\gamma_4(\L)=\frac{3}{32 \pi^2}g^4 \log \frac{\L^2}{p_0^2},\;\;\;\;\;
\nonumber\\
&&\nonumber\\
&&
\Delta^2_{\L,\L_0}=0\;\;\;\;\;\Delta^4_{\L,\L_0}= -\frac{g^4}{6\pi^3}
\frac{\left[\sqrt{(p_1+p_2)^2}-\sqrt{(\bar{p}_1+\bar{p}_2)^2}\right]+ [2\to3]
+[2\to4]}{\L}
\;,\nonumber\\
&&\nonumber\\
&&
\Gamma^{2n}_{\L,\L_0} = \frac{(-1)^{n+1}(2n)!}{2^n n (n-2) 32 \pi^2}
\frac{g^{2n}}{\L^{2n-4}},\;\;\;\;(n> 2).
\eeqra
The vanishing of $\gamma_2$ and $\Delta^2$ at one-loop is due to the momentum
independence of the tadpole. At two-loop they
get non-zero contributions which make them scale as $g^4 \log \L^2$ and 
$g^4 (p^2-p_0^2)^2/\L^2$ respectively.

The proof of perturbative UV renormalizability proceeds as follows \cite{bdm};
\begin{itemize}
\item[{ i)}] for any function $f^{2n}_{\L,\L_0}$ (could be any 
irrelevant vertex or momentum derivative of it)
define 
\beq
\label{MAX}
\left|f^{2n}\right|_\L \equiv \mathrm{Max}_{p_i^2<c \L^2}
\left|f^{2n}_{\L,\L_0}(p_1,\ldots,p_{2n})\right|\;,
\eeq
where $c$ is some $O(1)$ numerical constant;
\item[{ ii)}] at $l$-loop assume the following scalings
\beqra
\label{SCALINGS}
&&\gamma_2^{(l)}\sim\gamma_4^{(l)}\sim\left|\Delta^4_{(l)}\right|_\L =O(1)
\;,\;\;\;\;\gamma_3^{(l)}\sim \left|\Delta^2_{(l)}\right|_\L =O(\L^2)\;,
\nonumber\\
&&\nonumber\\
&&
\left|\partial^m\Gamma^{2n}_{(l)}\right|_\L
\sim \left|\partial^m\overline{\Gamma}^{2n}_{(l)}\right|_\L
=O(\L^{4-2n-m})\;,
\eeqra
where we have considered only the power-law behavior, since logarithms cannot 
change the power counting. The scaling behavior of 
$s_{(l)}(\L)$ can be read from 
eqs.~(\ref{KERNELSHARP},\ref{DELTA24},\ref{SCALINGS}),
\beq
\label{KERSCAL}
s_{(l)}(\L) =O(1)\;.
\eeq
Notice that none of these behaviors  depends on the loop order $l$.
\item[{ iii)}] using 
eq.~(\ref{INTEQIRREL}) we can maximize 
the $l+1$-loop contributions to the irrelevant vertices as follows 
\beqra
\label{maximD2}
&&\left|\Delta^2_{(l+1)}\right|_\L \lta \L^4 \int_{\L^2}^\infty d\lambda^2 
s_{(l-l^\prime)}(\lambda)
\left|\partial^4\Gamma^4_{(l^\prime)}\right|_\lambda\;,
\\
\label{maximD4}
&&\left|\Delta^4_{(l+1)}\right|_\L \lta \L^2 \int_{\L^2}^\infty d\lambda^2 
s_{(l-l^\prime)}(\lambda)
\left|\partial^2\overline{\Gamma}^6_{(l^\prime)}\right|_\lambda\;,
\\
\label{maximG2n}
&&\left|\Gamma^{2n}_{(l+1)}\right|_\L \lta \int_{\L^2}^\infty d\lambda^2 
s_{(l-l^\prime)}(\lambda)
\left|\overline{\Gamma}^{2n+2}_{(l^\prime)}\right|_\lambda\;,
\eeqra
where $l^\prime=0, \ldots, l$. In deriving
(\ref{maximD2}) and (\ref{maximD4}) we have used the fact that, 
due to the subtractions, the vertices of eq.~(\ref{SUBTR}) scale as
\beqra
&& \Delta\Gamma^4_{\L, \L_0}(q,p,-p,-q) \sim (p^2-p_0^2)^2
\frac{\partial^4\;\;}{\partial p^4} \Gamma^4_{\L, \L_0}(q,p,-p,-q)\,,
\nonumber\\
&&
\Delta\overline{\Gamma}^6_{\L, \L_0}(q,p_1,\ldots,p_4,-q)
\sim (P^2-\bar{P}^2)\frac{\partial^2\;\;}{\partial P^2}  
\overline{\Gamma}^6_{\L, \L_0}(q,p_1,\ldots,p_4,-q)\;,
\nonumber
\eeqra
where $P$ ($\bar{P}$) is a combination of $p_i$'s ($\bar{p}_i$'s).
Thanks to the scaling behaviors in eqs.~(\ref{SCALINGS},\ref{KERSCAL})
the $\L_0\to \infty$ limit can be taken, as the integrals are dominated by
the lower limit $\L$.
\item[{ iv)}]
the scalings assumed in eqs.~(\ref{SCALINGS},\ref{KERSCAL})  hold at one-loop, 
see eq.~(\ref{SCAL1LOOP}).
By using them
in eqs.~(\ref{maximD2}--\ref{maximG2n}), it is now straightforward to check
that the same-- \mbox{$l${\em-independent}}-- 
behaviors are obtained at $l+1$-loop. 
UV  
renormalizability is thus proved by induction  at any order.
\end{itemize}
In summary, the UV cut-off $\L_0$  can be removed, provided the three 
subtractions of eq.~(\ref{SUBTR})-- corresponding to the three 
renormalization conditions of eq.~(\ref{RCOND})-- are performed.
\subsection{IR finiteness}
In the previous subsection the $\L_0\to \infty$ limit at fixed-- and 
very large-- $\L$ was considered. In the massive-- and commutative! -- theory
the $\L\to 0$ limit can be taken with no particular care as the mass 
provides an effective IR cut-off regardless of the external momenta. 

The massless theory requires
a more careful study, which was also performed in \cite{bdm}. The statement to
be proved in this case is the finiteness in the $\L\to 0$ limit of any Green
function with no exceptional external momenta, where a couple of external 
momenta $p_i$ and $p_j$ is said to be exceptional if $p_i+p_j= O(\L)$. 
The proof proceeds again by induction, however it is technically complicated 
by the fact that the RHS's of the RG equations involve Green functions 
with a couple of exceptional momenta $q$ and $-q$. Thus, by iteration, 
Green functions with any number of pairs of exceptional momenta are involved,
and one has to consider the IR behaviors of all these. We do not need
here to give all the details, which can be found in \cite{bdm}. For future
comparison with the non-commutative case, we just recall the main feature of 
the commutative theory allowing IR finiteness being realized
order by order in the  loop expansion, namely the scaling behavior of 
the $l$-loop kernel,
\beq
\label{SIR}
s_{(l)}(\L)=O((\log\L)^{l-1})\;.
\eeq
As we will see in sect.~\ref{IR}, the mild logarithmic divergence above
is turned into a power-law in the non-commutative one, more and more divergent
as the loop order increases. As a consequence, any Green function is IR 
divergent at a sufficiently high order in the perturbative expansion, and
the latter has to be properly reorganized in order to obtain finite results.
\section{The non-commutative Wilsonian flow}
\label{NCWRG}
The RG equations (\ref{RGE2p},\ref{RGEnp})  hold for the 
non-commutative case as well.
In the Wilsonian framework, the non-commutativity of the theory is 
completely encoded in a 
different identification of the relevant vertex $\gamma_4(\L)$. Indeed,
we now write the four-point function as
\beq
\label{D4NC}
\Gamma^{4}_{\L, \L_0}(p_1,\ldots,p_{4})=
h(p_1,\ldots,p_4)\left[
\gamma_4(\L) +\Delta^4_{\L, \L_0}(p_1,\ldots,
p_{4})\right]\;,
\eeq
where the oscillatory function $h(p_1,\ldots,p_4)$ has been defined in
eq.~(\ref{FRULES}) and $\Delta^4_{\L, \L_0}(\bar{p}_1,\ldots,\bar{p}_{4})=0$.
The definitions of the other two relevant vertices $\gamma_2(\L)$ and 
$\gamma_3(\L)$ are the same as in eq.~(\ref{DELTA24}).
The renormalization conditions now are
\beq
\label{RCONDNC}
\Gamma_2(p^2=p_0^2)=m^2+p_0^2,\;\;\;\;\;\left.
\frac{d \Gamma_2(p^2)}{d p^2}\right|_{p^2=p_0^2}= 1,\;\;\;\;\;
\Gamma_4(\{\bar{p}_i\})= g^2 h(\bar{p}_1,\ldots,\bar{p}_4)\;,
\eeq
corresponding to the same initial conditions for $\gamma_{2,3}(\L)$ and the 
new $\gamma_4(\L)$ as those in eq.~(\ref{BCREL}).
At one-loop we now have the following contributions for $\gamma_2(\L)$ 
and $\gamma_3(\L)$ (for $m^2 \ll \mathrm{min}\{\L^2,1/\tilde{p_0}^2\}$)
\beqra
\label{REL1L}
&&\gamma_3(\L) = -\frac{g^2}{32 \pi^2}\left[\frac{2}{3} \L^2 + 
\frac{4}{3 \tilde{p}_0^2}(1-J_0(\L \tilde{p_0}))\right]\nonumber\\
&&\gamma_2(\L) = \frac{g^2}{24 \pi^2}\frac{1}{p_0^2\tilde{p_0}^2}
 \left[1 - J_0(\L \tilde{p_0})-\frac{\L \tilde{p_0}}{2}
J_1(\L \tilde{p_0})\right]\;\; .
\eeqra

Analogously, the irrelevant coupling $\Delta^2_{\L,\L_0}$, 
again at one-loop and neglecting the mass $m^2$, is
\beqra
\label{D1L}
&&\Delta^2_{\L,\L_0}=\frac{g^2}{24 \pi^2} \left\{\frac{J_0(\L \tilde{p})-
J_0(\L_0 \tilde{p})}{\tilde{p}^2}-
\frac{J_0(\L \tilde{p_0})-
J_0(\L_0 \tilde{p_0})}{\tilde{p_0}^2}
\right.\nonumber\\
&&\;\;\;\;\;\;\;\;\;\;\;\;\;\left. \frac{p^2-p_0^2}{p_0^2\tilde{p_0}^2}\left[
J_0(\L \tilde{p_0})-J_0(\L_0 \tilde{p_0})
+\frac{\L \tilde{p_0}}{2}J_1(\L \tilde{p_0})-
\frac{\L_0 \tilde{p_0}}{2}J_1(\L_0 \tilde{p_0})\right]\right\}.
\eeqra
Using the asymptotic behaviors of the Bessel functions, 
$J_0(x)\sim 1-x^2/4+x^4/64$, $J_1(x)\sim x/2-x^3/16$ for $x \ll 1$, 
 we can check the commutative limit of  eqs.~(\ref{REL1L}),~(\ref{D1L}). Indeed,
$\Theta^{\mu\nu}\to 0$, implies $\tilde{p}, \tilde{p_0}
\to 0$ for any $\L$, $\L_0$. In this case the results of eq.~(\ref{SCAL1LOOP})
are correctly reproduced.

We now consider $\gamma_4(\L)$ and $\Delta^4_{\L,\L_0}$: 
we do not have an analytical expression for any value 
of $\L,\L_0$ and $\bar{p}_i$: 
the relevant integrals are discussed in Appendix A. Taking $\L$ large 
($\L^2\gg p_0^2 \gg m^2$) we have: 
\beqra
\label{REL2L}
\gamma_4(\L) &&= \frac{g^4}{48\pi^2} \log \frac{\L^2}{p_0^2}-
\frac{g^4}{144\pi^2}\frac{1}{h(\bar{p}_1,\bar{p}_2,\bar{p}_3,\bar{p}_{4})}
\left[\cos (\bar{p}_{12})\cos (\bar{p}_{34})
\log\left(\frac{p_0^2(\tilde{\bar{p}}_1+\tilde{\bar{p}}_2)^2}{4}\right)\right.
\nonumber\\
&&\left. +\frac{1}{2}\cos (\bar{p}_{12})\cos (\bar{p}_{34})
\sum_i\log\frac{(p_0^2\tilde{\bar{p}}_i^2)}{4}+
\frac{1}{4}\cos(\bar{p}_{12}+\bar{p}_{34})
\log\left(\frac{p_0^2(\tilde{\bar{p}}_1+\tilde{\bar{p}}_4)^2}{4}\right)\right.
\nonumber\\
&&\left. + \frac{1}{4}\cos(\bar{p}_{12}-\bar{p}_{34})
\log\left(\frac{p_0^2(\tilde{\bar{p}}_1+\tilde{\bar{p}}_3)^2}{4}\right)+
[2\to3]+[2\to4]\right]\,.
\eeqra
We see a very 
different behavior between the planar contribution ({\it i.e.} the first 
logarithmic term), that is similar to the 
commutative case, and the non-planar one, producing a complicate structure 
that depends logarithmically on the non-commutative scale. 
On the other hand, if we take the commutative limit at fixed $\L$
the non-planar part develops 
exactly the factor $\ds \frac{7g^4}{96\pi^2} \log \frac{\L^2}{p_0^2}$, thus
reproducing the result of eq.~(\ref{SCAL1LOOP}).

Concerning $\Delta^4_{\L,\L_0}$ we consider the $\L_0\to \infty$ 
limit and evaluate the scaling behavior at large $\L$:  
\beqra
\label{SCAL4}
\Delta^4_{\L,\infty}=&& \frac{g^4}{27\pi^2}
\frac{1}{h(\bar{p}_1,\bar{p}_2,\bar{p}_3,\bar{p}_{4})}
\left[\cos (\bar{p}_{12})\cos (\bar{p}_{34})
\frac{\sqrt{(p_1+p_2)^2}-\sqrt{(\bar{p}_1+\bar{p}_2)^2}}{\L}\right]\nonumber\\
&& + [2\to3]+[2\to4]\,\,\,+O(1/\L^2).
\eeqra
The non-planar contribution is heavily suppressed in the above limit: one can 
nevertheless check that at ${\it finite}$ $\L_0$, the commutative limit 
is correctly reproduced.

We are now ready to extend the proof of UV renormalizability described in 
the previous section to the non-commutative case.

\subsection{UV renormalizability of the non-commutative theory}
We start by defining the non-commutative scale $M_{NC}$ as
\beq
\label{Massa}
M_{NC}\equiv (6 \pi^2 \Tr {\bf A})^{-1/4}\,,
\eeq
where the entries of the matrix ${\bf A}$ are given by
$A_{\mu\nu}=-\Theta_{\mu\rho}\Theta_{\rho\nu}$.

In order to prove UV renormalizability we assume the following relations 
between the mass scales of the theory,
\beq
\label{HIE}
\frac{M_{NC}^2}{\L_0}\ll p,\;p_0,\;M_{NC}\ll \L\ll\L_0\;,
\eeq 
where $p$ is a generic external momentum. The two extremal
inequalities above are not restrictive, as we are interested in the 
$\L_0\to \infty$ limit. In particular, the lower bound on $p$ and $p_0$ 
ensures that
$\Delta^2_{\L\L_0}$ and $\Delta^4_{\L\L_0}$ 
are finite and $\L_0$-independent in that limit.
The condition on $\L$ has been assumed, as before, 
in order to have a simple power-counting. Notice that 
 at this stage we assume that $\L$ is also larger than the
non-commutative scale $M_{NC}$. The practical consequence of this 
is to cut-off the contributions from the non-planar graphs to the irrelevant
vertices, so that all the terms containing explicit $\T^{\mu\nu}$ dependence, 
like  those  seen in $\Delta^2_{\L\L_0}$ and $\Delta^4_{\L\L_0}$, are suppressed.

Going back to sect.~\ref{RENCOMM} we will now maximize the irrelevant vertices 
with a slight modification of the definition of $|f^{2n}|_\L$, {\em i.e.}
\beq
\left|f^{2n}\right|_\L \equiv \mathrm{Max}_{(M_{NC}^2/\L_0)<p_i<c \L}
\left|f^{2n}_{\L,\L_0}(p_1,\ldots,p_{2n})\right|\;,
\eeq
with $c$ again some $O(1)$ numerical constant.

At one-loop the scalings in eq.~(\ref{SCALINGS}) still hold, as can be 
checked directly from eqs.~(\ref{REL1L}, \ref{D1L}). It remains to be proved 
that, assuming they hold at $l$-loops, they still hold at $l+1$-loops once
the integrals in eqs.~(\ref{maximD2}, \ref{maximD4}, \ref{maximG2n}) 
are performed. The only thing we have to check is the behavior of 
$s_{(l)}(\L)$. From eq.~(\ref{KERNELSHARP}) we know that it contains the
self-energy of eq.~(\ref{DELTA24}) computed at external momentum $\sim \L$. 
Then, using the scaling laws of eq.~(\ref{SCALINGS}) we have
\beq
\label{ScalUV}
s_{(l)}(\L) =O(1)\;,
\eeq
as in (\ref{KERSCAL}).
As a consequence, the maximizing integrals are the same as for the commutative
case, and this is enough to prove perturbative UV renormalizability for the 
non-commutative case.
\subsection{The IR regime and the need of a resummation}
\label{IR}
When the IR regime comes under scrutiny things change considerably, the 
reason being that the scaling (\ref{ScalUV}) does not hold any more
in the $M_{NC} \gg \L \to 0$ regime. Indeed, the dominant IR behavior
of the one-loop self-energy in the IR is given by
\beq
\Sigma^{(1)}_{\L,\L_0}(p)
= \frac{g^2}{24 \pi^2 }\left(\frac{1}{\tilde{p}^2}
-\frac{\L^2}{4}
+\cdots\right)\;.
\eeq 
Recalling that the function $s(\L)$ contains $\Sigma_{\L,\L_0}(\L)$ (see
eq.~(\ref{KERNELSHARP})), we have, at $l$-loop
\beq
s_{(l)}(\L) \sim \left[\frac{M_{NC}^4g^2}{\L^2 (\L^2+m^2)}\right]^l +
\cdots\;,
\eeq
where the dots represent less IR-divergent terms. The most IR-singular 
contributions come from the so-called `daisy' diagrams, 
{\em i.e.} multiple insertions of (non-planar) one-loop tadpoles. It 
is then clear that any Green function, even those without exceptional 
external momenta, is IR divergent at a sufficiently high order in the 
loop expansion. For instance,
the two-point function diverges quadratically in the IR at two-loop
for the massless 
theory and logarithmically at three-loop for the massive one, more 
tadpole insertions giving more and more IR-divergent behaviors. 

Looking at the exact form of the RG kernel, eq.~(\ref{KERNELSHARP}),
we realize how a solution can be found. Since 
$\Sigma_{\L\L_0}(\L)$
appears in the denominator, it is clear that the full
equations are indeed better behaved in the IR than any approximation
to them computed at a finite order in $g^2$. Actually, since the
self-energy acts as an effective mass exploding  in the  $\L\to 0$ limit,
they are even better behaved than those for the commutative massive theory! 

It appears then plausible
that the IR pathologies might 
just be an artifact of the perturbative expansion,
which could disappear if this is properly reorganized.
To see that this is indeed the case, one can split the full
two-point function as
\beqra
\label{SPLIT}
\ds 
\Gamma^{(2)}_{\L\L_0}(p) = &&(p^2+m^2)
K(p;\L,\L_0)^{-1}+\Sigma^{\mathrm 1-loop}_{\mathrm LO}(p)\nonumber\\
&& + 
\Delta\Sigma_{\L,\L_0}(p)\,,
\eeqra
where $\Sigma_{\mathrm LO}=g^2/24 \pi^2 \tilde{p}^2$ is the leading IR 
divergence at one-loop. 
Eq.~(\ref{SPLIT})
 defines the new `tree-level'
 propagator as 
\beq
\label{RESPROP}
\frac{1}{(p^2+m^2)
K_{\L,\L_0}(p)^{-1}+\Sigma_{\mathrm LO}(p)}\;,
\eeq
and the new tree-level 
kernel as eq.~(\ref{KERNELSHARP}) with $s_{(0)}(\L)= 
[1+\Sigma_{\mathrm LO}(\L)/(\L^2+m^2)]^{-2}$.
We can then define a new perturbative procedure to solve the RG equations, in
which each iteration adds a new loop with these propagators in the internal 
lines. 

Again, we have to prove that any Green function without exceptional external 
momenta is finite in the $\L\to 0$ limit. 
With respect to the commutative case, we have now to extend the definition of 
exceptional  momenta including not only the case $p_i+p_j=O(\L)$, but
also $\ds \tilde{p}_i+\tilde{p}_j=O\left(\frac{\L}{M_{NC}^2}\right)$.
The demonstration that the new perturbative expansion is free from IR 
divergences is now obvious.
The tree-level propagator
vanishes as $M_{NC}^{-4} \L^2$ if it carries an exceptional momentum. 
Then, given
a Green function and a loop order, the higher the number of external 
exceptional momenta, the lower the degree of IR divergence. So the most 
worrisome functions would be those with all external momenta of the 
non-exceptional type. 
At one-loop in the resummed expansion, the latter are IR finite. 
Assuming they are finite at 
$l$-loop, they are also so at $l+1$-loop, since the loop integration induces
a factor at most 
\[
d\lambda \frac{\lambda^3}{\lambda^2+m^2} s_{(0)}(\lambda) \sim d\lambda 
(\lambda^2+m^2) \lambda^7 M_{NC}^{-8}\;
\]
in the IR.

The behavior of the four-point function does not spoil the above conclusion.
Indeed, in the extreme IR limit (and for $\L_0\to \infty$), one gets 
\newpage
\beqra
\label{REL3L}
\lim_{\L\to\infty}
\Delta^4_{\L\L_0} &&= \frac{g^4}{144\pi^3} \log \frac{(p_1+p_2)^2}{p_0^2}-
\frac{g^4}{144\pi^2}
\left[\log\left(\frac{(p_1+p_2)^2(\tilde{p}_1+\tilde{p}_2)^2}
{(\bar{p}_1+\bar{p}_2)^2(\tilde{\bar{p}}_1+\tilde{\bar{p}}_2)^2}\right)\right.
\nonumber\\
&&\left. +\frac{1}{2}\sum_i\log\left(\frac{p_i^2\tilde{p}_i^2}
{\bar{p}_i^2\tilde{\bar{p}}_i^2}\right)+
\frac{1}{4}\log\left(\frac{(p_1+p_2)^2(\tilde{p}_1+\tilde{p}_4)^2}
{(\bar{p}_1+\bar{p}_2)^2(\tilde{\bar{p}}_1+\tilde{\bar{p}}_4)^2}\right)
\right.
\nonumber\\
&&\left. + \frac{1}{4}\log\left(\frac{(p_1+p_2)^2(\tilde{p}_1+\tilde{p}_3)^2}
{(\bar{p}_1+\bar{p}_2)^2(\tilde{\bar{p}}_1+\tilde{\bar{p}}_3)^2}\right)\right]
+[2\to3]+[2\to4].
\eeqra
We see, therefore, that the four-point function develops only 
${\it logarithmically}$ divergent IR singularities. When appearing into 
higher order graphs, they are made harmless by the presence of a resummed 
propagator carrying the same loop momentum.

As an application of this resummation, we will review in the next section 
our recent calculation of the 
next-to-leading  correction to the two-point function \cite{gp1}. 
\subsection{Hard non-commutative loop resummation}
In this section, we will abandon for a while the RG framework to 
formulate the resummation of IR divergences discussed in the previous section 
in a more common diagrammatic language. The need of a 
resummation has been realized by different people, and discussed for instance
in \cite{minwa,resu}, although not in a systematic way. 

The procedure outlined in the previous section
can be rephrased by adding and
subtracting the term
\beq
\label{CT}
\frac{g^2}{48 \pi^2} \int 
\frac{d^4p}{(2\pi)^4} \phi(p)\frac{1}{\tilde{p}^2}\phi(-p)
\eeq
from the tree-level lagrangian, so as to get the resummed propagator 
(\ref{RESPROP}) provided
the new two-point `interaction' in eq.~(\ref{CT}) is 
also taken into account, in very
close analogy to what is done for the resummation of IR divergences in 
finite temperature scalar theory
\cite{parwani}.

The interactions of the resummed theory give the 
Feynman rules in Fig.~1.
\begin{figure}[htb]
\begin{center}
\epsfxsize=2.4 in \epsfbox{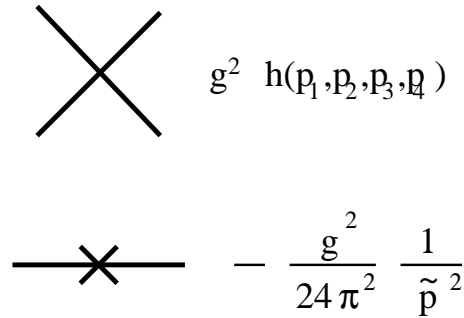}
\end{center}
\caption{The interaction vertices of the resummed theory}
\end{figure}
\noindent

Now we can compute the next-to-leading order corrections to the 
self-energy, which are given by the two diagrams in Fig.~2, where the
resummed propagator runs into the loop (of course also the graph with the UV 
counterterms has to be included, which is not shown in the figure).
\begin{figure}[htb]
\leavevmode
\begin{center}
\epsfxsize=2.4 in \epsfbox{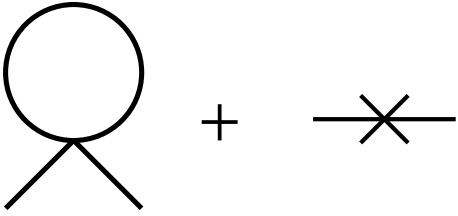}
\end{center}
\caption{The next-to-leading order contributions to the self-energy}
\end{figure}
\noindent
The tadpole diagram in the resummed theory gives
\beqra 
\Delta\Sigma(p)&& =
\frac{g^2}{6} \int\frac{d^4 q}{(2 \pi)^4} \frac{1}{q^2+m^2+\frac{g^2}{24\pi^2}
\frac{1}{\tilde{q}^2}} \left[2 + \cos(q\cdot \tilde{p})\right] \nonumber\\
&&-
\frac{g^2}{24\pi^2}
\frac{1}{\tilde{p}^2} + ``{\mathrm UV c.t.}''\,. 
\eeqra
In the UV, the integral has the 
same structure as for the non-resummed theory, with a quadratically divergent
contribution from the `planar' diagrams and a finite one from the `non-planar'
ones, giving the $1/\tilde{p}^2$ term which is exactly cancelled by the 
new two-point interaction of the resummed theory.
In the IR, the planar and non-planar contributions sum up.
By writing 
\[\tilde{q}^2=\frac{1}{4}\Tr {\bf A} q^2 + q\cdot{\bf B} \cdot q\,,
\] where $B_{\mu\nu}$ is a 
traceless symmetric matrix. The symmetry of the integrand in the IR regime,
 selects  the $q^2$ term as the dominant contribution.

In the massless case ($m=0$) we find the following contribution from the 
`planar' graph
\beq
\label{MLP}
\Delta\Sigma^{m=0}_{\mathrm pl}(p)= -\frac{g^3}{96 \pi} M_{NC}^2 +O(g^5)\,,
\eeq
whereas from the `non-planar' one we get
 \beqra
\label{MLNP}
\Delta\Sigma^{m=0}_{\mathrm npl}(p)=&&\ds -\frac{g^3}{192 \pi} M_{NC}^2 
\nonumber \\
&& - \frac{g^4 M_{NC}^4 \tilde{p}^2}{1536 \pi^2} \left[
\log \frac{g^2 M_{NC}^4 \tilde{p}^4}{256}- {\mathrm const}\right]\,
\eeqra
for $M_{NC} \tilde{p} \ll 1$ and 
\[
\Delta\Sigma^{m=0}_{\mathrm npl}(p)= O\left(\frac{g}{M_{NC}^4 \tilde{p}^6}
\right)\,,
\]
for $M_{NC} \tilde{p} \gg 1$.
In the massive case we get (planar + non-planar)
\beqra
\label{MASS}
\Delta\Sigma&&=\frac{g^2}{8 \pi^2}\left[m^2 \log m^2 -
\frac{g^2}{4}\frac{M_{NC}^4}{m^2}\right.\nonumber\\
\ds
&&\ds -\left.\frac{g^4}{8}\frac{M_{NC}^8}{m^6}
\left(\log\frac{g^4 M_{NC}^8}{m^8} +3\right) + O(g^6)\right]\,.
\eeqra
As one could expect, the non-analyticity in the coupling $g^2$ emerges 
at lower order in the massless case (where we find a $O((g^2)^{3/2})$ 
correction) compared to the massive one ($g^6 \log g^4$). This reflects
the fact that, in ordinary perturbation theory, 
the self-energy is IR divergent at $O(g^4)$ in the former 
case and  at $O(g^6)$ in the latter.

In computing the next-to-next-to-leading order in the resummed perturbative 
expansion one must
consistently take into account the two-point interaction in (\ref{CT}). 
Indeed, 
the two-loop graph for the resummed $m=0$ theory with one 
non-planar tadpole insertion (first graph in Fig.~3) 
gives a contribution of $O(g^3 M_{NC}^2)$, the 
same as the corrections computed above. It is only when the graph containing
the 
two-point interaction is added that the whole correction comes
out $O(g^5 M_{NC}^2)$.
\begin{figure}[htb]
\leavevmode
\begin{center}
\epsfxsize=2. in \epsfbox{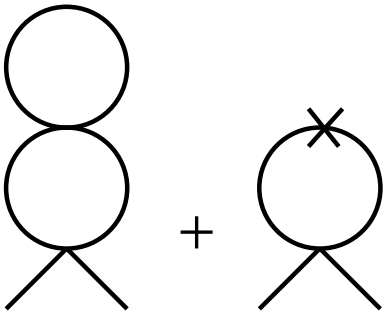}
\end{center}
\caption{The higher order corrections}
\end{figure}
\noindent
The $O(g^4)$ corrections that one gets at two-loop, coming from UV loop
momenta, cannot modify the $O(g^4 \log g^2)$ term in eq. (\ref{MLNP}).

The corrections computed above are really `perturbatively small' compared 
to the leading two-point function
 $p^2+g^2/24 \pi^2 \tilde{p}^2$ in any range of the momentum $p$. Indeed, for
large enough  momenta, the $\Delta \Sigma$ correction dominates over the 
$g^2/\tilde{p}^2$ term, but in that regime the tree-level $p^2$ term is 
leading. On the other hand in the IR the opposite happens, with $\Delta \Sigma$
never dominating over $g^2/\tilde{p}^2$.
As a consequence, no 
 tachyonic behavior can be induced by the next-to-leading order 
corrections.

\section{IR/UV connection and the breakdown of the Wilsonian picture}
\label{PRED}
In the previous sections we were interested in the UV 
renormalizability of the theory, so we considered the $\L_0\to\infty$ limit. 
In this
section we take a different perspective. We define a high-energy 
theory at $\L_0$, and ask how it looks like at lower and lower energies and 
larger and larger distances, {\it i.e} as $\L$ and the external momenta are 
lowered, while $\L_0$ is kept fixed. In the commutative theory, the well known 
Wilsonian picture holds:  by properly redefining the three
relevant couplings any sensitivity to the high scale vanishes as some power 
of $\L/\L_0$ or $p/\L_0$, $p$ being some generic external momentum. Then, the
irrelevant couplings at some scale $\L$  depend on $\gamma_{2,3,4}(\L)$ and
on $\L$ itself, modulo the above mentioned power law corrections.

In the non-commutative theory the Wilsonian picture breaks down. It can be
seen explicitly by looking at the one-loop expressions for the irrelevant
couplings in eq.~(\ref{D1L}) and recalling the limiting expressions for the
Bessel functions in eq.~(\ref{LIMITS}). As long as the external momentum $p$ 
and the subtraction point $p_0$ are such that 
$\tilde{p}\L_0,\:\tilde{p}_0 \L_0 \gg 1$, the $\L_0$-dependence  in 
$\Delta^{2,4}_{\L,\L_0}$ is exponentially suppressed\footnote{Indeed,
the oscillating Bessel functions of eq.~\ref{D1L} vanish only in an average 
sense. This is just a technical point, due to our choice of a sharp momentum 
cut-off. The choice of a smooth (exponential or power-law) cut-off does
not alter the discussion in the previous sections and gives rise to
 non-oscillating functions, exponentially vanishing for
$\tilde{p}\L_0,\:\tilde{p}_0 \L_0 \gg 1$.}.
In the previous section,
we used this fact to take the $\L_0\to\infty$ limit. 
On the other hand, if we lower the external momenta so that 
$\tilde{p}\L_0 \lta 1$ then we get
\beq
\Delta^{2}_{\L,\L_0}(p) = \frac{g^2}{96 \pi^2} \L_0^2+\cdots\;,
\eeq
and analogously $\Delta^{4}_{\L,\L_0}\propto \log(\L/\L_0)$. Thus, the
Wilsonian picture is spoiled, as the theory at 
large distances becomes  sensitive to the small distance scale 
$\L_0$. 

The origin of this IR/UV connection lies in the UV behavior of non-planar
diagrams. They are cut-off by the smaller between $\L_0$ and $1/\tilde{p}$, 
then it is clear that the $\L_0$ dependence of UV divergent diagrams is 
screened for $p \gg M_{NC}^2/ \L_0$, and shows up when $p$ is lowered beyond
this threshold.
If the non-commutative scale $M_{NC}$ and
the scale of the  high energy theory $\L_0$, are well separated, 
two different situations may arise:
\begin{itemize}
\item if $\L_0 \ll M_{NC} $ the non-planar diagrams behave as the planar
ones for $p,\,\L<\L_0$ ({\it i.e.} they are cut-off by $\L_0$ in the UV). 
The Wilsonian flow reduces to  that of the commutative theory from $\L=\L_0$ down
to $\L=0$ and the $\L_0$ sensitivity is suppressed by powers of $\L/\L_0$ or
$p/\L_0$. Therefore the usual Wilsonian picture holds, since in this energy
range we are basically in the commutative regime.
\item if  $\L_0 \gg M_{NC}$, that is, the high-energy theory is defined
in the deeply non-commutative region, a Wilsonian picture holds in the
range $M_{NC}^2/\L_0 \ll p,\,p_0,\,\L \ll \L_0$, where the RG flow is truly
non-commutative (planar and non-planar contributions evolve differently). As
we have discussed above, the picture breaks down if we try to lower
$p$ beyond $M_{NC}^2/\L_0$.
\end{itemize}
\begin{figure}[htb]
\leavevmode
\begin{center}
\epsfxsize=5.8 in \epsfbox{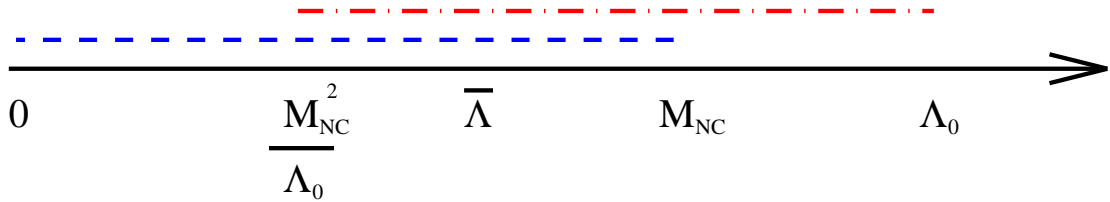}
\end{center}
\caption{The two overlapping ranges in which a Wilsonian picture holds}
\end{figure}
\noindent
\subsection{Matching two Wilsonian flows}
From the above analysis we learn that the Wilsonian picture does not hold 
straightforwardly from a very high scale $\L=\L_0 \gg M_{NC} $ down 
to $\L=0$. Nevertheless, it may be 
partially recovered as a two-steps procedure, in which the high-energy theory
is matched to a low-energy one at a intermediate scale
 $\bar{\L}$ 
such that $M_{NC}^2/\L_0 \ll \bar{\L} \ll M_{NC}$. $\bar{\L}$
lies in the region in which  the Wilsonian picture holds for both the 
high-energy and the low-energy RG flows. Then, once the $\L_0$-dependence is 
absorbed in the relevant couplings of the high-energy
theory, it does not enter the matching conditions to the 
low-energy one at $\bar{\L}$ . 
Since the RG flow for $\L,\,p\lta \bar{\L}$ is that of a
commutative theory, the low-energy theory knows about non-commutativity only 
via its boundary conditions, {\it i.e.} the matching conditions at $\bar{\L}$.
In the following we illustrate this procedure explicitly at one-loop.

By giving boundary conditions at $\L_0$, the relevant couplings of the 
high-energy theory are
\beqra
\label{RELH}
&&\gamma_3^h(\L) =\gamma_3^h(\L_0) 
+\frac{g^2}{32 \pi^2}\left[\frac{2}{3} \L_0^2 - 
\frac{4}{3 \tilde{p}_0^2}J_0(\L_0 \tilde{p_0}) - (\L_0\to \L) 
\right]\nonumber\\
&&\gamma_2^h(\L) =\gamma_2^h(\L_0)+ 
\frac{g^2}{24 \pi^2}\frac{1}{p_0^2\tilde{p_0}^2}
 \left[J_0(\L_0 \tilde{p_0})+\frac{\L_0 \tilde{p_0}}{2}
J_1(\L_0 \tilde{p_0}) - (\L_0\to \L) \right]\;\;\;\;\nonumber\\
&&\gamma_4^h(\L) = \gamma_4^h(\L_0) +
 \frac{g^4}{48\pi^2} \log \frac{\L^2}{\L_0^2} +
\cdots\;\;\;\;\;\;\;\;(m^2\ll p_0^2,\;\L^2)\,,
\eeqra
where dots are exponentially suppressed terms, 
while for the low-energy theory they are given by the same expressions
with the substitutions 
\beq
\label{HTOL}
``h'' \to ``l'', \;\;\;\;\;\;\;\;\; \L_0 \to \bar{\L}.
\eeq
 The $\L_0$
dependence in (\ref{RELH}) can be absorbed in the boundary conditions
$\gamma_{2,3,4}^h(\bar{\L})$, and
the matching conditions for the relevant couplings are trivially given by
\beq
\gamma_{2,3,4}^h(\bar{\L}) =\gamma_{2,3,4}^l(\bar{\L}).
\eeq
Besides the relevant couplings, we have also to impose matching conditions
on the two irrelevant ones $\Delta^2_{\L,\L_0}$ and $\Delta^4_{\L,\L_0}$.
We start by discussing $\Delta^2_{\L,\L_0}$. For the high-energy 
theory, $\Delta^{2,h}_{\L,\L_0}(p)$ is given
by eq.~(\ref{D1L}), while for the low-energy one,  
$\Delta^{2,l}_{\L,\bar{\L}}(p)$, we have to perform the same substitution as 
in eq.~(\ref{HTOL}) and to add the initial condition
$\Delta^{2,l}_{\bar{\L},\bar{\L}}(p)$(the initial conditions  
at $\L_0$ for 
the irrelevant couplings of the high energy theory are taken to be zero, as in 
eq.~(\ref{D1L})).

To fix the matching conditions we first take 
$\L=\bar{\L}$ and set the external 
momenta at some $P= O(\bar{\L}) $. Equating the irrelevant couplings for
the two theories we get,
\beqra
\label{MATCHIRR}
\Delta^{2,l}_{\bar{\L},\bar{\L}}(P)&&
= \Delta^{2,h}_{\bar{\L},\L_0}(P)\nonumber\\
&&=
\frac{g^2}{24 \pi^2} \left\{\frac{J_0(\bar{\L} 
\tilde{P})}{\tilde{P}^2}-
\frac{J_0(\bar{\L} \tilde{p_0})}{\tilde{p_0}^2} + 
\frac{{P}^2-p_0^2}{p_0^2 \tilde{p_0}^2}\left[
J_0(\bar{\L} \tilde{p_0})
+\frac{\bar{\L} \tilde{p_0}}{2}J_1(\bar{\L} \tilde{p_0})\right]
\right\} \nonumber\\
&&\;\;\;\;\;+ 
({\mathrm{exp.\; suppressed \; terms\; in\; }}\L_0),
\eeqra
where we have neglected the terms in $\L_0$ since,
 due to or choice for the external momenta $P$
 (and for $p_0$, which we take in the same range), 
they are exponentially suppressed. 
Now, we choose the boundary conditions for momenta $p\le O(P)$ by using
the same expression (\ref{MATCHIRR}), {\em i.e} we neglect
 the $\L_0$-terms {\it for any value of the external momenta},
\beqra
\label{IRRLE}
&&\Delta^{2,l}_{\L,\bar{\L}}(p) = \Delta^{2,l}_{\bar{\L},\bar{\L}}(p) 
\nonumber\\
&& +
\frac{g^2}{24 \pi^2} \left\{\frac{J_0(\L 
\tilde{{p}})}{\tilde{{p}}^2}-
\frac{J_0(\L \tilde{p_0})}{\tilde{p_0}^2} + 
\frac{{p}^2-p_0^2}{p_0^2 \tilde{p_0}^2}\left[
J_0(\L \tilde{p_0})
+\frac{\L \tilde{p_0}}{2}J_1(\L \tilde{p_0})\right]
 - (\L\to \bar{\L)}
\right\} 
\nonumber\\
&& \;\;\;\; =\frac{g^2}{24 \pi^2}\left[\frac{1}{\tilde{p}^2} - 
\frac{1}{\tilde{p_0}^2} + \frac{p^2-p_0^2}{p_0^2\tilde{p}_0^2}\right] +
O({\L}^4\tilde{p}^2,\, \L^4\tilde{p}_0^2),
\eeqra
where $\Delta^{2,l}_{\bar{\L},\bar{\L}}(p)$ is given by 
eq.~(\ref{MATCHIRR})) for 
$P\to p$.
Notice  that 
the contribution to $\Delta^{2,l}_{\L,\bar{\L}}(p)$ comes mainly from its 
boundary condition at $\bar{\L}$. The running from
$\bar{\L}$ downwards contributes with $O(\bar{\L}^4 p^4/M_{NC}^8)$ suppressed
terms, consistent with the fact that $\Delta^{2}_{\L,\bar{\L}}(p)$ 
is zero at one-loop in the commutative theory.

Analogously for the 
four-point irrelevant vertex we impose
\beq
\Delta^{4,l}_{\bar{\L},\bar{\L}}(P_i)=\Delta^{4,h}_{\bar{\L},\L_0}(P_i)\,,
\eeq
and, neglecting the $\L_0$ terms for lower external momenta, we get
(in the $\L\to 0$ limit)
\beqra
\lim_{\L\to 0}\Delta^{4,l}_{\bar{\L},\bar{\L}}(p_i)\simeq &&
\frac{g^4}{32\pi^2}\log\left(\frac{(p_1+p_2)^2}{p_0^2}\right)-
\frac{g^4}{144\pi^2}\left[\log\left(\frac{p_0^2(\tilde{\bar{p}}_1
+\tilde{\bar{p}}_2)^2}{4}\right)+\frac{1}{2}
\sum_i\log\frac{(p_0^2\tilde{\bar{p}}_i^2)}{4}\right.\nonumber\\
&&\left. +\frac{1}{4}\log\left(\frac{p_0^2(\tilde{\bar{p}}_1
+\tilde{\bar{p}}_4)^2}{4}\right)+ 
\frac{1}{4}\log\left(\frac{p_0^2(\tilde{\bar{p}}_1+\tilde{\bar{p}}_3)^2}{4}
\right)
\right]\nonumber\\
&&+(2\to 3)+(2\to 4).
\eeqra
The first logarithm above is just the 1-loop four-point function of the
commutative theory, the rest is the contribution coming from the 
boundary condition at $\bar{\L}$.

From the low-energy point of view, the information that the original 
high energy theory 
is non-commutative is encoded in the boundary conditions at $\bar{\L}$  
for the three
relevant couplings plus those for the two irrelevant ones
$\Delta^2_{\L,\bar{\L}}$ $\Delta^4_{\L,\bar{\L}}$. 
This is to be contrasted with the usual Wilsonian 
picture, where the dependence on the 
boundary conditions of all the irrelevant couplings vanishes  at 
low energies as some  powers of $p/\bar{\L}$.
Indeed, this is exactly what makes a theory predictive, in that it depends 
only on a finite number of boundary conditions. 
In the non-commutative case, the new scale $M_{NC}$ gives rise to
terms like $1/\tilde{p}^2$ and $\log{\tilde{p}}$ which do not decouple
when $M_{NC} \gg \bar{\L} \gg p \to 0$. Fortunately 
 such terms enter only $\Delta^2_{\L,\bar{\L}}$ and $\Delta^4_{\L,\bar{\L}}$,
 but not $\Gamma_{2n}$
for $n \ge 3$, as can be checked dimensionally. 
Thus, the boundary conditions for the latter couplings become
more and more irrelevant as $p, \L \to 0$  and 
the low-energy theory is still predictive --and $\L_0$-independent-- 
although it needs
two extra boundary conditions compared to the commutative case.

From the point of view of the low-energy observer, the two extra boundary 
conditions might come from some high-energy degrees of freedom, as discussed in
\cite{minwa}. Also in that picture, once the extra degrees of freedom
are integrated out, a commutative theory with the bizarre 
propagator $(p^2+g^2/24 \pi^2 \tilde{p}^2)^{-1}$ is obtained. However, in the 
case of the massive theory, the 
reproduction of the logarithmic $\tilde{p}$ behavior   by the same means 
turns out to be not so straightforward. 

\section{Summary and outlook}
\label{CONC}
The Wilsonian RG equations (\ref{INTEQIRREL}) exhibit a 
remarkable momentum ordering; a given irrelevant coupling evaluated at
cut-off
$\L$ receives contributions only from loop momenta $\lambda \ge \L$. 
In this paper 
we have used this basic property in order to split the analysis of the 
perturbative behavior of the non-commutative scalar theory in two steps.
First,
by taking $\L$ much larger than any physical mass scale, we have
inspected
the UV sector of the theory. In this regime, if the external momenta are
all 
$\gg M_{NC}^2/\L_0$, the contribution of the non-planar diagrams is
damped
by the non-commutative phases while that of the planar ones is the same
as for
the commutative theory. The proof of perturbative renormalizability at
any
order in perturbation theory is then 
just a straightforward generalization of that given for the commutative
case in
\cite{polchinski, bdm}.

Then we turned to the IR sector by lowering $\L$ towards the physical
limit, 
$\L\to 0$. In this regime the well-known IR/UV connection spoils
perturbation 
theory completely, as IR divergences appear in the contributions to any
Green
function. The IR divergences can be completely resummed in a way
analogous
to what is customarily done in finite temperature field theory, {\it
i.e.}
by using a resummed propagator and introducing a corresponding
counterterm in the interaction lagrangian. The resummation procedure
does not
change the UV sector of the theory, so that the previous discussion of
UV 
renormalizability holds unaltered. In the resummed theory we were able
to 
compute the next-to-leading corrections to the two-point functions,
which 
exhibit a non-analytic dependence on the coupling constant.

Finally, we discussed what survives in the non-commutative 
case of the usual Wilsonian picture, {\it i.e.}
insensitivity of the theory at long distance to the short distance
behavior
after proper redefinition of the relevant couplings.
At first sight, the IR/UV connection seems to spoil completely this
picture. 
Indeed, since {\em in the resummed theory} 
 the IR/UV connection affects only the two- and four- point 
functions, the situation is less dramatic. The high-energy theory can be
matched to a low-energy one which in the IR limit  
has a purely commutative RG flow and
knows about non-commutativity only via the boundary conditions of the 
two- and four- point functions. If the scale at which the high-energy
theory 
is defined, $\L_0$, is  well above the non-commutative scale, the
low-energy 
theory sensitivity to $\L_0$ is exponentially suppressed.

The present study opens a series of questions. The first one is the
extension
of this analysis to gauge theories.  The main problem in that context 
is  gauge 
invariance, which is broken by the introduction of a momentum cut-off
and can
be recovered only in the physical limit $\L\to 0$, $\L_0 \to \infty$. 
At finite values of the cut-off the theory satisfies modified Ward
identities
which were discussed for the commutative case in \cite{bdmgauge}. In
that paper
it was  shown how gauge invariance can be controlled order by 
order in perturbation theory in such a way as to be recovered in
the physical limit.  
The extension of those results to the non-commutative case
requires close scrutiny, the main complication being the need of a
resummation in order to get a finite $\L\to 0$ limit. It is not
presently 
clear to us how such a resummation can be performed. 
It is evident that the simple 
resummation of the self-energy in the propagator is not a gauge
invariant 
operation. Again, the example of thermal field theories might be of help
for 
us. In the pure QCD case, Ward identities connect n to n+1-point
functions, 
so that not only the gluon self-energy has to be resummed, but 
all n-gluon functions. The program was achieved by Braaten and 
Pisarski and leads to 
to the well known Hard Thermal Loop effective action \cite{bp}. It would
be 
very interesting to understand if an analogous result can be obtained in
the 
non-commutative case.

Another interesting issue is that of phase transition and the critical 
regime. Indeed some analysis at one-loop  have been already presented in 
ref.~ \cite{pt2}, but we argue that the actual behavior should be
quite different from that discussed in that paper. 
The point is that, being higher loops IR divergent,
the study of the critical regime can be consistently performed
only in the resummed theory, which has the propagator  
$(p^2 + m^2 + g^2/24\pi^2\tilde{p}^2)^{-1} $. 
So, the long distance $p\to 0$ limit 
is dominated by the $g^2/24\pi^2\tilde{p}^2$ term and turns out to be
insensitive to the sign of $m^2$, which usually determines whether the
vacuum 
breaks or not the symmetry. Thus we conclude that there are no phase 
transitions in these theories, at least of the common type. In the RG
language,
the flow shuts-off well before the mass can be probed 
( if $g^2 M_{NC}^4/m^4 \gg1$) and the critical exponents are drastically 
changed with respect to those of the massless commutative theory.
This agrees with the evidences for a non-homogeneous phase 
discussed in ref.~\cite{pt1}.

%\begin{figure}[h]

%\leavevmode

%\hspace{2.5cm} \epsfbox{fig3d.ps}

%\end{figure}

%\noindent

%\setcounter{section}{0}
%\Alph{section}
\appendix
\section{Appendix}
\label{Appy}
We have at one-loop that
\beq
\gamma_4(\Lambda)=I_4^{(p)}+I_4^{(np)},
\label{a1}
\eeq
where the integrals defining $\gamma_4(\Lambda)$ are
\beqra
I_4^{(p)}&&=\frac{2g^4}{3}\int\frac{d^4q}{(2\pi)^4} 
\frac{\Theta(\L^2-q^2) \Theta(\bar{p}^2-2\bar{p}q) }{(q^2+m^2)[(q-\bar p)^2+m^2]}
\nonumber\\
I_4^{(np)}&&=\frac{g^4}{9}\frac{1}{h(\bar p_1,\bar p_2,\bar p_3, \bar p_4)}
\int\frac{d^4q}{(2\pi)^4} 
\frac{\Theta(\L^2-q^2) \Theta(\bar{p}^2-2\bar{p}q) }{(q^2+m^2)[(q-\bar p)^2+m^2]}
F(q,\bar p_i)\nonumber\\
&&+(\bar{p}_2\to \bar{p}_3)+(\bar{p}_2\to \bar{p}_4).
\label{a2}
\eeqra
$F(q,\bar p_i)$ is defined as
\beqra
F(q,\bar p_i)&&=2\cos\bar p_{12}\cos\bar p_{34}\cos (\bar p\cdot\tilde{q})
+\cos\bar p_{34}\left[\cos \left(\bar p_{12}+\bar p_1\cdot\tilde{q}\right)+
\cos \left(\bar p_{21}+\bar p_2\cdot\tilde{q}\right)\right]
\nonumber\\
&&+\cos\bar p_{21}\left[\cos \left(\bar p_{43}+\bar p_4\cdot\tilde{q}\right)
+\cos \left(\bar p_{34}+\bar p_3\cdot\tilde{q}\right)\right]
\nonumber\\
&&+\frac{1}{2}\cos \left[\bar p_{12}+\bar p_{34}+\left(\bar{p}_1+\bar{p}_4-
\bar{p}_2-\bar{p}_3\right)\cdot\tilde{q}\right]\nonumber\\
&&+\frac{1}{2}\cos \left[p_{12}-p_{34}+\left(\bar{p}_1-\bar{p}_4-\bar{p}_2
+\bar{p}_3
\right)\cdot\tilde{q}\right].
\label{a3}
\eeqra
The planar contribution $I_4^{(p)}$ is evaluated for $\L\gg p_0\gg m$
\beq
\frac{g^2}{12\pi^3}\int_{\frac{p_0}{2}}^{\L}dq^2\int^{1}_{0}dx
\sqrt{1-x^2}\frac{1}{q^2+p_0^2+2\bar p qx}\simeq
\frac{1}{48\pi^2}\log\frac{\L^2}{p_0^2}.
\label{a4}
\eeq
The generic integral involved in the non-planar part is
\beq
\int\frac{d^4q}{(2\pi)^4} 
\frac{\Theta(\L^2-q^2) \Theta(\bar{p}^2-2\bar{p}q) }{(q^2+m^2)((q-\bar p)^2+m^2)}
\cos(a+b\cdot\tilde{q}).
\label{a5}
\eeq
Assuming $\L$ very large
\beqra
\simeq\frac{1}{4\pi^3}&&\left[  \cos a \int_0^1 dx\sqrt{1-x^2}
\left({\rm ci}(\L\tilde b x)-{\rm ci}
(\frac{p_0\tilde b x}{2})\right)+\right.\nonumber\\
&& \left.\sin a \int_0^1 dx\sqrt{1-x^2}\left({\rm si}(\L\tilde b x)-{\rm si}
(\frac{p_0\tilde b x}{2})\right)\right],
\label{a6}
\eeqra
where we have introduced the sine-integral and the cosine-integral functions. 
As $\L\to\infty$ both ${\rm ci}(\L\tilde b x)$ and ${\rm si}(\L \tilde b x)$ 
goes to zero. 
Taking instead $\L$ finite and $\tilde b\to 0$, and using 
\beqra
&&{\rm ci}(\L\tilde b x)-{\rm ci}(\frac{p_0\tilde b x}{2})\simeq \log\frac{2\L}{p_0}
 +O(\tilde b),\nonumber\\
&&{\rm si}(\L\tilde b x)-{\rm si}(\frac{p_0\tilde b x}{2})\simeq O(\tilde b),
\label{a7}
\eeqra
we obtain that the leading contribution is commutative-like 
$$\simeq\cos a \log\frac{\L^2}{p_0^2}$$, 
while as $\L\to \infty$ the dominant term is
\beq
-\frac{1}{4\pi^3}\cos a \int_0^1 dx\sqrt{1-x^2}{\rm ci}
(\frac{p_0\tilde b x}{2}).
\label{a8}
\eeq
For $p_0\tilde b \gg 1$ the contribution is very small while as $p_0\tilde b \leq 1$ 
we can expand the integral using eq. (\ref{a7}) and to recover 
eq. (\ref{REL2L}) where the 
explicit form of $F(q,\bar p_i)$ has been used.

Concerning $\Delta^4_{\L\L_0}$ we have a similar separation (at one-loop):
\beq
\Delta^4_{\L\L_0}=\Delta^4_{p}+\Delta^4_{np},
\eeq
where for $\L_0\to \infty$ we have
\beqra
\Delta^4_{(p)}&&=\frac{2g^4}{9}\int\frac{d^4q}{(2\pi)^4} 
\left[\frac{\cos p_{12}\cos p_{34}}{h(p_i)}
\frac{\Theta(q^2-\L) \Theta({p}^2-2{p}q) }{(q^2+m^2)((q-p)^2+m^2)}
-(p\to \bar p)\right]_{p=p_1+p_2}\nonumber\\
&&+(p_2\to p_3)+(p_2\to p_4)\nonumber\\
&&=\frac{g^4}{18\pi^3}
\int dq~q \int_{0}^{1}dx\sqrt{1-x^2}\left[ 
\frac{\cos p_{12}\cos p_{34}}{h(p_i)(q^2+p^2-2pqx)}-
(p\to \bar{p})\right]_{p=p_1+p_2}
\nonumber\\
&&+(p_2\to p_3)+(p_2\to p_4).
\label{a9}
\eeqra
For large $\L$ we can expand the denominators retaining the linear terms and 
obtaining therefore the first term in eq. (\ref{SCAL4}). The non-planar part is 
instead suppressed in the same limit: the typical contributions have the form
\beq
\int_{\L}^{+\infty}\frac{dq}{q}\int_{-1}^{1}dx\sqrt{1-x^2}\cos (\tilde{b}qx)
\simeq \int_{\L}^{+\infty} \frac{dq}{q}\frac{J_1(\tilde{b}q)}{\tilde{b}q}.
\eeq

\end{document}